\documentclass{eas}
\usepackage{graphicx}
\def\mnras{MNRAS}

\def\aap{A\&A}
\def\nat{Nature}
\def\apj{ApJ}
\def\apjl{ApJ Letters}

\def\mp{m\ind{p}}\def\mg{m\ind{g}}
\def\kp{k\ind{p}}\def\kg{k\ind{g}}\def\kc{k\ind{c}}

\newcommand{\ind}[1]{_{\mathrm{#1}}}
\newcommand{\diff}{\mathrm{d}}

\def\Kepler{\emph{Kepler}}

\def\numax{\nu\ind{max}}\def\nmax{n\ind{max}}

\newcommand{\np}{n}

\def\Dnu{\Delta\nu}

\def\Dnuobs{\Delta\nu\ind{obs}}
\def\Dnuas{\Delta\nu\ind{as}}

\def\epsobs{\varepsilon\ind{obs}}
\def\epsg{\varepsilon\ind{g}}
\def\coup{q}
\def\Tg{\Delta\Pi_1}

\begin{document}

\title{Sounding stellar cores with mixed modes}
\runningtitle{Red giant asteroseismology}
\author{Beno{\^\i}t Mosser, K\'evin Belkacem, Mathieu Vrard}\address{LESIA, CNRS, Universit\'e Pierre et Marie Curie,
Universit\'e Denis Diderot, Observatoire de Paris, 92195 Meudon
cedex, France; \email{benoit.mosser@obspm.fr}}
%
%
\begin{abstract}
The space-borne missions CoRoT and \emph{Kepler} have opened a new
era in stellar physics, especially for evolved stars, with precise
asteroseismic measurements that help determine precise stellar
parameters and perform ensemble astero\-seismology. This paper
deals with the quality of the information that we can retrieve
from the oscillations. It focusses on the conditions for obtaining
the most accurate measurement of the radial and non-radial
oscillation patterns. This accuracy is a prerequisite for making
the best with asteroseismic data. From radial modes, we derive
proxies of the stellar mass and radii with an unprecedented
accuracy for field stars. For dozens of subgiants and thousands of
red giants, the identification of mixed modes (corresponding to
gravity waves propagating in the core coupled to pressure waves
propagating in the envelope) indicates unambiguously their
evolutionary status. As probes of the stellar core, these mixed
modes also reveal the internal differential rotation and show the
spinning down of the core rotation of stars ascending the red
giant branch. A toy model of the coupling of waves constructing
mixed modes is exposed, for illustrating many of their features.
\end{abstract}
\maketitle

\section{Introduction}

Red giant seismology is one of the most beautiful unexpected gift
of the space missions CoRoT and \Kepler. When the CoRoT mission
was planned, red giants were seen as unavoidable targets because
of their brightness. Their observation has however confirmed that
they show solar-like oscillations: as in the solar case, turbulent
convection in the uppermost atmosphere excite pressure
oscillations.

Pressure oscillation modes detected in low-mass stars at different
evolution stages, all along the red giant branch (RGB), allow us
to perform ensemble asteroseismology. From such study, expressed
by scaling relations, we can derive global stellar properties and
get keys for deciphering stellar evolution. Mixed modes, absent in
the Sun, have been discovered in red giants (Beck et al. 2011).
Since they result from the coupling of pressure waves propagating
in the envelope with gravity waves propagating in the inner
radiative regions, they can be used to directly probe the stellar
core, from which they extract unique information on the ongoing
nuclear rotation (Bedding et al. 2011, Mosser et al. 2011a), or on
the inner rotation rate (Beck et al. 2012, Deheuvels et al. 2012,
Mosser et al. 2012c).

A recent review on red giant oscillations can be found in
\cite{2013EPJWC..4303003M}. Results obtained in red giant
seismology are discussed in a companion paper (Mosser et al.
2013b), with an emphasis on the seismic scaling relations and on
red giant interior structure. This proceedings paper deals with
tricky points related to the data analysis. In Section
\ref{radial}, we discuss different ways to describe and measure
the radial-mode oscillation pattern and explain the conditions
providing the most precise measurement of the large separation. In
Section \ref{dipole}, we focus on the dipole mixed-mode pattern
and present a toy model used to explain the conditions under which
mixed modes are observed. This models helps understand the
so-called depressed mixed modes observed in RGB stars, or the
large variety of observable gravity-dominated mixed modes.

\section{Radial modes\label{radial}}

We aim to obtain an efficient description of the radial-mode
oscillation pattern, for the most precise measurement of their
frequencies. A first step for determining the characteristics of
the oscillation pattern consists in the identification of radial
modes and in the measurement of the large separation $\Dnuobs$, as
it appears as frequency difference between observable modes. The
radial oscillation pattern is, at first-order approximation (e.g.,
Tassoul 1980), $\nu_{n,0} \simeq ( n+\epsobs (\Dnuobs)) \, \Dnuobs
$.
This form derives from an asymptotic analysis and is therefore
valid for large radial orders $n$.

In this Section, we first discuss different methods for measuring
$\Dnuobs$. This large separation is measured around the frequency
$\numax$ of maximum oscillation signal, in fact in non-asymptotic
conditions. This explains that the measured value $\Dnuobs$ must
be distinguished from its asymptotic counterpart. Then, we
investigate the different meanings of the large separation, and
also the difference between local and global measurements of
$\Dnuobs$. Finally, we discuss the meaning of the offset parameter
$\epsobs$.

\subsection{Envelope autocorrection function}

Different methods have been proposed for measuring the large
frequency separation of solar-like oscillations. All these methods
have been exhaustively compared, in both main-sequence and red
giant regimes (e.g., Verner et al.~2011., Hekker et al.~2011).
Among all methods, the autocorrelation of the time series presents
different advantages (Roxburgh \& Vorontsov 2006, Mosser \&
Appourchaux 2009). Computationally, the method is rapid since it
is based on Fourier transforms. For computing the Fourier spectrum
of the original time series, a Lomb-Scargle periodogram is
required. Then, all following operations are fast Fourier
transforms of the Fourier spectrum. It allows an automated
process, efficient for all large separations in the whole range of
solar-like oscillations (from 0.1 to 200$\,\mu$Hz). It especially
works at very low frequency, where, contrary to other methods, the
relative poor frequency resolution is not an issue. Dedicated
filters can be used to compute the spectrum of the filtered
Fourier spectrum of the time series. These filters can be adapted
to many situations, from the most efficient global measurement of
the mean value of the large separation at $\numax$ to its rapid
variations with frequency (Mosser 2010). The method can also be
used for measuring other frequency separations than the large
frequency separation. It has been used successfully to measure the
mixed mode spacings (Mosser et al.~2011b) or the
rotational splittings (Mosser et al.~2012c).\\

Many other methods are used for computing the large separation.
Since $\Dnuobs$ represents the period of the comb-like structure
of the oscillation spectrum, its value can be inferred from the
autocorrelation of the spectrum. We however note that this
autocorrelation, contrary to the autocorrelation of the time
series, does not benefit from the rapid calculation provided by
the Wiener-Khinchin theorem. For space-borne observations
benefitting from a very high duty cycle, this method appears to be
less efficient than the autocorrelation of the times series.

\subsection{Different meanings of the large separations}

The concept of large separation has different meanings, depending
on the context. Observationally, the large separation $\Dnuobs$ is
measured from the difference between consecutive radial mode
frequencies. The local or global nature of the value is discussed
in the next paragraph. Mosser et al. (2013) have shown that
$\Dnuobs$ is significantly different from the asymptotic value
$\Dnuas$, which is another important acceptation. It corresponds
to the value introduced by theoretical asymptotic methods (Tassoul
1980) and it scales as the inverse of the acoustic diameter of the
star. The large separation intervenes in scaling relations
(Belkacem et al. 2013), under the hypothesis that is very close to
the dynamical frequency $\nu_0$ that scales with
$\sqrt{\mathcal{G}
M/R^3}$.\\

The values $\Dnuobs$, $\Dnuas$ and $\nu_0$ are close to each
other, but different. The measurement of the large separation
$\Dnuobs$ must try to recover, if possible, $\Dnuas$ and $\nu_0$.
As shown by \cite{2013A&A...550A.126M}, linking $\Dnuas$ to
$\Dnuobs$ is possible if and only if the second-order asymptotic
expansion is considered. Then, one has to account for the fact
that the large separation is measured in non-asymptotic
conditions. The curvature seen in the spectrum is the signature of
this second-order term. The relation between $\Dnuobs$ and
$\Dnuas$ writes $\Dnuas = (1 + a (\Dnuobs)) \, \Dnuobs$
The correction term is maximum in the red giant regime; it
corresponds there to a uniform correction, about 3.8\,\%.
Confusion is often made between $\Dnuas$ and $\nu_0$. Since
scaling relations are extensively used for translating the global
seismic parameters into seismic proxies of the mass and radius, it
is necessary to avoid this confusion. Modeling is therefore
useful, as done recently by \cite{2013arXiv1307.3132B} who show
that, in most cases, the observed large separation is closer to
$\nu_0$ than the asymptotic value $\Dnuas$.

\subsection{Global or local measurements of the large separations?\label{local-global}}

The different methods used for measuring the large separation
provide either a local or a global value of the frequency
difference between radial modes. For a local measurement, only 2
or 3 radial orders around $\numax$ are considered: $\Dnuobs \simeq
\Dnu_n =  \nu_{n+1,0} - \nu_{n,0}$, with the radial frequencies
$\nu_{n,0}$ and $\nu_{n+1,0}$ close to $\numax$. For a global
measurement, $\Dnuobs$ is a weighted mean value calculated in a
broad frequency range around $\numax$.

Evil is in the detail: the difference between the local and global
measurement is small, but with a systematic component.
\cite{2012A&A...541A..51K} have shown that this small difference
can be used for distinguishing red giants in the red clump or on
the RGB. This difference is due to glitches (Miglio et al.~2010)
that depend on the evolutionary stage. Therefore, measuring both
the local and global large separations is very useful. This
implies that the modulation due to the glitches is not negligible:
it perturbs the measurement of the large separation, in a way that
is not related to any of the global properties of the asymptotic
large separation or dynamical frequencies. Therefore, using a
global value of $\Dnuobs$ is certainly the best way to get rid of
the glitches that significantly modulate the large separation.
Furthermore, the work by Mosser et al.~(2013) has shown that the
oscillation pattern obeys closely the second-order asymptotic
oscillation pattern. This implies that a global large separation
will provide a better measurement than a local value, for allowing
the translation into the asymptotic value.

The last step of the demonstration of the advantages of the global
measurement is exposed later (Section~\ref{offset-global}). One
needs first to present a method permitting an efficient global
measure.

\subsection{The universal oscillation pattern}

For red giants, an efficient measure is made easy by the large
homology of the stellar interiors. \cite{2011A&A...525L...9M} have
shown that interior structure homology of red giants translates
into homology of their oscillation spectrum. As a consequence, the
large separation is enough to parameterize the low-degree spectrum
\begin{equation}\label{univ}
    \nu_{n,\ell} = \left[ n + \epsobs + {\alpha\over 2} (n-\nmax)^2
    + d_{0\ell} \right] \ \Dnuobs,
\end{equation}
where the offset $\epsobs$, the curvature $\alpha$, and the
relative separations $d_{0\ell}$ are functions of the observed
large separation $\Dnuobs$. The curvature term was already
suspected in asteroseismology with ground-based observations with
a limited frequency resolution, as for instance the Procyon
observations reported by \cite{2008A&A...478..197M}. This term is
necessary to account for the fact that the large separation is
measured around $\numax$, in non-asymptotic conditions; in fact,
it accounts for the gradient in frequency separations ($\Dnu_n =
(1+\alpha(n-\nmax)) \Dnuobs$, according to the derivation of
Eq.~\ref{univ}). The term $\nmax$ is defined by
\begin{equation}\label{nmax}
    \nmax = \numax/\Dnuobs - \epsobs
\end{equation}
so that $\Dnuobs$ is the large separation measured at the
frequency $\numax$. Such a formalism ensures  the most precise
translation from the measurement at $\numax$ to the asymptotic
value. The term $\nmax$ can be seen as the radial order at
$\numax$. However, contrary to a real radial order, $\nmax$ is not
an integer.

In practice, measuring $\Dnuobs$ with the universal pattern is
derived from the fit of the observed radial frequencies with a
pattern based on Eq.~\ref{univ}. Quadrupole modes are in fact
fitted too, but not dipole modes, since they show a mixed
character and have a more complex structure. The use of the
universal pattern has proven to be efficient, with the following
properties: i) it accounts for the frequency separation gradient;
ii) the fitting process helps reduce the realization noise due to
the fact that the modes are short lived; iii) it also lowers the
influence of the modulation due glitches discussed above; iv) it
provides a measure which is not affected by the mixed-mode
pattern.

\subsection{Accuracy of the universal pattern}

A large part of the accuracy of the universal pattern is due to
the introduction of second-order terms. In Eq.~\ref{univ}, the
second-order term is expressed by the curvature $\alpha$, which
helps for fitting the radial ridge in a large frequency range. If
the second-order term were not taken into account, the error for
identifying low or high radial-order radial modes, assuming a
perfect match at $\numax$, could be as large as $0.1\,\Dnuobs$,
which is similar to the separation between quadrupole and radial
modes. Consequently, the measure of the large separation is less
precise when the curvature is omitted.

The accuracy of the large separations determined from the
universal pattern depends on the validity of the relation found
for $\epsobs (\Dnuobs)$  in the asymptotic expansion. From
\cite{2013A&A...550A.126M}, it is clear that this term is not a
surface term: its value is the direct signature of the measurement
of the large separation around $\numax$, in non-asymptotic
conditions\footnote{In the solar case, the difference between
$\epsobs \simeq 1.5$ and $\varepsilon\ind{as}= 0.25$ is entirely
explained by the non-asymptotic value of the commonly used large
separation.}. We can translate an uncertainty in the $\varepsilon
(\Dnuobs)$ relation into an uncertainty in $\Dnuobs$. In fact, the
derivation of the first-order relation of the radial mode writes:
\begin{equation}\label{erreurs}
    \diff\epsobs = - (n+\epsobs)\ {\diff \Dnuobs \over \Dnuobs} .
\end{equation}
Assuming a perfect match at $\numax$ between the observed spectrum
and the universal pattern, but a relative displacement $\eta$ (in
unit $\Dnuobs$) at the extremities of the energy excess envelope,
we infer a relative precision of the measurement of the large
separation about $2\eta / N$, where $N$ is the total number of
modes. For sake of simplicity, we take $N \simeq \nmax$ (Mosser et
al. 2010).

We can estimate $\eta$ in two ways, from the amplitude of the
glitches measured in the red giant oscillation spectrum (Miglio et
al. 2010), or by comparison with the relative distance $d_{20}$
between quadrupole and radial modes. A typical value of the glitch
amplitude (Vrard et al., in preparation) is of the order of 2\,\%.
This gives a relative precision in the measurement of the large
separation of the order of 0.5\,\% (or equivalently 0.02\,$\mu$Hz)
at the red clump, in agreement with exhaustive tests and
comparisons of different methods performed by
\cite{2011A&A...525A.131H}. The comparison between $\eta$ and
$d_{20}$ provides a similar estimate, which quantitatively does
not significantly change with $\Dnuobs$.

These results demonstrate the high accuracy of the method, even at
low signal-to-ration or with short time series (Hekker et al.
2012).

\subsection{$\epsobs$ offset and evolutionary status\label{offset-global}}

We can now interpret the influence of the glitches mentioned in
Section~\ref{local-global}.  When the large separation is measured
locally, the term $\epsobs$ is not simply a function of the large
separation, but depends also on the evolutionary status: stars on
the RGB or in the red clump have not similar $\epsobs$. The
(small) difference disappears when the large separation is
measured globally, in a large frequency range, as demonstrated by
\cite{2012A&A...541A..51K}. This indicates that the $\epsobs
(\Dnuobs)$ relation does not depend on the evolutionary status
when $\Dnuobs$ is measured globally.

We can investigate the signature of a systematic difference in the
$\epsobs$ relation between the clump and RGB stars. This
difference is about 0.15, according to \cite{2012A&A...541A..51K}.
According to Eq.~\ref{erreurs}, such a difference in $\epsobs$
translates into a 1.5\,\% difference in $\Dnuobs$ at the red
clump. This implies then a relative shift of $0.06\,\Dnuobs$ for
radial modes at radial orders close to $\nmax\pm 4\Dnuobs$. This
shift represents one half of the relative separation between the
radial and quadrupole modes. It is definitely not observed when
one compare RGB and clump stars oscillation patterns, proving that
there is no significant systematic dependance of $\epsobs$ with
the evolutionary status. It also underlines that the most accurate
values of the large separations are measured globally and not
locally. As stated above, the local measurement of $\Dnuobs$ is
very valuable for measuring, by comparison with the global value,
the evolutionary status; however, a local value is less precise
for later use in scaling relations (see Mosser et al. 2013b).

\section{Dipole modes\label{dipole}}

The radial modes (together with the quadrupole modes) have proven
to be highly useful for identifying the red giant oscillation
spectrum, from early RGB (Deheuvels et al.~2012) to red giants in
the upper RGB and AGB (Mosser et al.~2013b). The dipole modes are
then most useful to carry information from the stellar core (Beck
et al.~2011, Bedding et al.~2011, Mont\'alban et al.~2013).

\subsection{Mixed pressure and gravity modes\label{mixed}}

Dipole modes result from the coupling of pressure \emph{waves}
excited in the upper stellar envelope with gravity \emph{waves}
propagating only in the inner radiative region. In red giants, the
latter region approximately corresponds to the core (except in the
helium burning region of the core of red clump stars) and its
surrounding hydrogen-burning shell.

One often reads that mixed modes result from the coupling of
gravity \emph{modes} propagating in the core with pressure
\emph{modes} propagating in the envelope. This shortcut, even if
fully clear, does not help understand the physics of mixed modes.
Apart from the fact that modes, as standing waves, do not
propagate, the coupling of modes is an insecure concept, which
does not help understand the physics and the conditions on the
\emph{phase} of the different waves for constructing the modes.
It is more relevant to imagine a wave excited by
turbulent convection in the upper envelope, propagating in the
star as a pressure wave (with pressure acting as a restoring
force) until it reaches the Lamb frequency, evanescent beyond this
limit, and propagating again but as a gravity wave (with buoyancy
acting as a restoring force) in the radiative core. The coupling
conditions ensure that various frequencies, and not only the
frequency of the pure pressure mode that would exist in absence of
the radiative cavity, may efficiently couple. The resonant
conditions then yield to the mixed modes. In red giants, we
observed up to $\mathcal{N}+1$ mixed modes per $\Dnu$-wide
frequency range. They result from  one pressure wave and
$\mathcal{N} = \numax^2 \Tg / \Dnu$ gravity waves, where $\Tg$ is
the asymptotic period spacing of gravity modes (Mosser et
al.~2012b).

\subsection{Asymptotic expansion}

An asymptotic expansion has been used by
\cite{2012A&A...540A.143M} to estimate the mixed-mode frequencies.
It is based on the pressure-mode asymptotic pattern, and makes
profit of the formalism developed by \cite{1989nos..book.....U}.
It writes as an implicit equation:
\begin{equation}
\nu = \nu_{\np,\ell=1} + {\Dnu \over \pi} \arctan %
\left[%
 \coup \tan \pi \left( {1 \over \Tg \nu} -
 \epsg
\right) \right] . \label{implicite2}
\end{equation}
The frequency $\nu_{\np,\ell=1}$ correspond to a pure pressure
mode. It relatively differs from the radial mode by a term
$d_{01}$ close to 0.52 (Mosser et al.~2011a). The phase coupling
factor $\coup$ is discussed below. The offset $\epsg$ is supposed
to be 0.

The accuracy of the asymptotic expansion of mixed modes has not
yet been directly tested. However, numerous indirect tests have
been made, which indicate the ability of the asymptotic expansion
to accurately fit the data:

- It is based, for the pressure part, on the universal oscillation
pattern, which is fully operational for radial modes. For the
gravity component, it takes the Tassoul asymptotic formalism of
gravity modes into account. Since,  the gravity radial orders are
quite high ($|n\ind{g}|$ in the range [50 -- 500] except in the
very early RGB), the asymptotic expansion is supposed to apply
precisely.

- \emph{All} high signal-to-noise spectra showing mixed modes
could be fitted accurately (including, if necessary, rotational
splittings).

- The method provides valuable proxies for the mixed modes that
are then successfully fitted by modelling (Jiang et al. 2011, Di
Mauro et al.~2011, Deheuvels et al.~2012). In complex cases with a
high density of mixed modes, the asymptotic expansion is the only
way to disentangle the different components of the mixed modes and
to provide the identification of the modes.

Equation \ref{implicite2} introduces a coupling coefficient $q$,
as in the original expansion developed by
\cite{1989nos..book.....U}. This coupling must not be considered
as an energetic coupling, related to the influence on the
evanescent regions. On the contrary, it accounts for the relative
weights of the pressure and gravity waves in the mixed modes. If
the mode is equally p or g (this occurs when the density  of pure
pressure modes per frequency interval is equal to the density of
gravity modes), the phase coupling factor $q$ is close to 1. This
does not occur in the red giant regime, but for subgiants (Benomar
et al.~2013). On the contrary, when the density of pure p modes is
much lower, then the coupling factor is small, and significantly
smaller for RGB stars compared to clump stars.

\begin{figure}[t!]
\centering
\includegraphics[width=7.8cm]{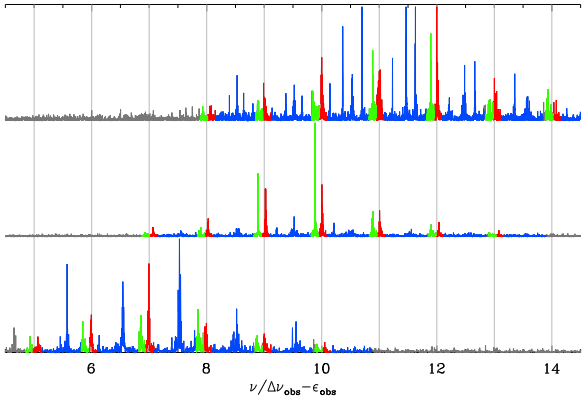}
\caption{Red giant oscillation spectra, as a function of the
normalized frequency;
$\ell=0$, 1, and 2 modes are plotted in red, blue, and green,
respectively. \textbf{Top:} typical pattern with a large number of
gravity-dominated mixed modes. \textbf{Middle:} dipole modes are
depressed. \textbf{Bottom:} dipole modes are mostly
pressure-dominated. \label{diff_visi}}
\end{figure}

\begin{figure}[t!]
\centering
\includegraphics[width=7.8cm]{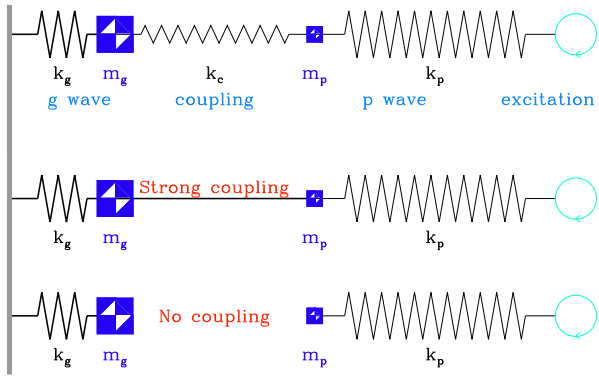}
\caption{Toy model with two masses and three springs,
representative of the coupling of gravity and pressure waves. {\bf
Top:} the mean case, with an intermediate coupling, helps explain
the gravity-dominated mixed modes. {\bf Middle:} very strong
coupling; this case provides an explanation of depressed mixed
modes. {\bf Bottom:} very weak coupling; in this case, only
pressure modes are observed. \label{ressort}}
\end{figure}

\subsection{Gravity dominated, pressure dominated, and depressed
dipole mixed modes}

The dipole modes observed in red giants show a large variety of
oscillation patterns (Fig.~\ref{diff_visi}). Except at low
$\numax$, all these patterns show a large number of
gravity-dominated mixed modes that follow the mixed-mode
asymptotic expansion (Eq.~\ref{implicite2}). The amplitudes of
these modes also show a large variety of cases: dipole mixed modes
have very low amplitudes for a family of stars identified by
\cite{2012A&A...537A..30M}; in other cases analyzed by
\cite{2012A&A...540A.143M}, red giants present a rich dipole
pattern with a large number of gravity-dominated mixed modes; in
other cases, the pattern is practically limited to the
pressure-dominated mixed modes. Apart from the modelling effort to
interpret such mixed modes (Dupret et al. 2009, for massive
stars), we propose a simple mechanical model for trying to
understand the different cases, including the depressed mixed
modes.

In this toy-model, we consider a chain of three springs and two
masses, as indicated in Fig.~\ref{ressort}. The stiffness and the
mass that, respectively, represent the p and g modes have to
fulfil the resonance condition: $\omega^2 = \kp / \mp = \kg /
\mg$. The masses $\mp$ and $\mg$ are representative of the
pressure and gravity mode inertia, respectively. This implies that
$\mg\gg\mp$, since the inertia of gravity modes is much higher
than the inertia of pressure modes. This means that, accordingly,
the stiffness $\kg$ of the gravity spring is much stronger than
the stiffness $\kp$ of the pressure spring. The stiffness of the
coupling spring is, as one can imagine, representative of the
coupling (note that this coupling has an energetic meaning,
contrary to the phase coupling factor discussed above). Finally,
the excitation mechanism occurs at the extremity of the pressure
spring, in order to mimic the free surface of the oscillation and
the location of the excitation in the uppermost level. The
opposite extremity of the gravity spring is fixed; it corresponds
to the barycenter of the star.

This toy model can be used to understand typical observations. We
first examine the most common case, where a large number of
gravity-dominated mixed modes are observed (Fig.~\ref{diff_visi}).
We then investigate two opposite (and extreme) cases corresponding
to the case where the amplitudes of gravity-dominated mixed modes
are severely damped, or to the case where only pressure-dominated
mixed modes are seen:

- The common case corresponds to a `moderate' coupling. In the
transitory regime, the coupling is low enough to ensure the
efficient excitation of pressure waves: the energy input from the
oscillation is not significantly perturbed by the leak through the
coupling. In a permanent regime, these (p-)waves have enough
energy to excite then gravity waves. Therefore, the model helps
understand the gravity dominated mixed modes. The stiffness $\kc$
of the coupling spring is low enough to allow significant
amplitudes of the p mode, but high enough to regulate the
frequency of this p mode by the much more stable gravity
oscillator. Depending on the frequency, the resulting modes are
dominated either by pressure or gravity.

- A tight coupling corresponds to a very high coupling stiffness.
In that case, we can consider that the masses $\mp$ and $\mg$ are
bound together. Since $\kg \gg \kp$, no significant movement is
expected: the excitation at the surface is not efficiently
transmitted to $\mp$ and $\mg$ since they are in practice tightly
attached by $\kg$ with the immobile center. The excitation is
inefficient: the waves excited at the surface cannot reach great
amplitudes since that they have to move a very high inertia. Even
the pressure-dominated modes cannot form. It is straightforward to
consider that the depressed mixed modes, with limited oscillation
amplitudes, correspond to this strong-coupling case. This might
correspond to a reduced region between the pressure and gravity
cavity. In other words, the Brunt-V\"ais\"al\"a (BV) and $\ell=1$
Lamb (L1) frequencies have close values, which are also close to
the frequency of the damped modes. If the BV frequency exceeds the
Lamb frequency, one may suppose that, with pressure and buoyancy
acting as restoring forces, the coupling will be extremely
efficient.

- An inefficient coupling is ensured by an absence of the coupling
spring. In that case, the excitation is able to move $\mp$ without
any reaction of $\mg$. This case corresponds to oscillation
spectra without important gravity dominated peaks. Such a coupling
may correspond to a wide region between the BV and L1 cavities. In
that case, gravity waves cannot be excited.


In this model, it is possible to suppose, as in the real case,
that the power density of the excitation only weakly depends on
frequency. All mixed modes associated to the same pressure mode
have then similar height in a power density spectrum. They just
differ by their lifetime. A condition to reach uniform height is
the observation duration, which has to be high enough, higher than
the mixed-mode lifetimes, in order to observe resolved modes.

Finally, this model helps understand the energy partition reported
by \cite{2012A&A...537A..30M}. The excitation taking place in the
uppermost envelope is transferred towards the stellar core by p
waves only. The total energy that could be affected to one single
p mode is shared between all mixed modes associated to this p
mode. This partition is efficient in most cases, when the coupling
between the waves in the two cavities is moderate. A high coupling
induces an important consequence: the dipole modes have a much
higher inertia than radial modes, so that their surface amplitudes
are totally damped.

\subsection{Rotation and mixed modes}

Rotational splittings observed in a handful of giants have first
exhibited the significant radial differential rotation in the red
giants interior (Beck et al.~2012, Deheuvels et al.~2012): the
inner rotation rate is significantly higher than the envelope
rotation rate. Then, the analysis of $\simeq\,300$ spectra by
\cite{2012A&A...540A.143M} has shown the spin-down of the mean
core rotation when red giant ascent the RGB. This spin-down is
more pronounced for more-evolved stars in the red clump.

\cite{2013A&A...549A..74M} have established seismic diagnostics
for transport of angular momentum in stars from the pre-main
sequence to the red-giant branch. They found that transport by
meridional circulation and shear turbulence yields far too high a
core rotation rate for red-giant models compared with
observations. Either turbulent viscosity is largely
underestimated, or a mechanism not included in current stellar
models of low mass stars is needed to slow down the core rotation.
\cite{2013A&A...549A..75G} have derived a theoretical framework
for understanding the properties of the observed rotational
splittings of red giant stars with slowly rotating cores. They
were then able to describe the morphology of the rotational
splittings, under the hypothesis of linear splittings. In fact,
the mean core rotation dominates the splittings, even for pressure
dominated mixed modes. As a consequence, they provided theoretical
support for using a Lorentzian profile to measure the observed
splittings of red giant stars.  \cite{2013A&A...554A..80O} have
shown the complexity of the splittings when rotation cannot be
considered as a perturbation. When rotational splitting are higher
than  twice the frequency separation between consecutive dipolar
modes, non-perturbative computations are necessary. Each family of
modes with different azimuthal symmetry $m$ has to be considered
separately. In case of rapid core rotation, the differences
between the period spacings associated with a given azimuthal
order $m$ constitutes a promising guideline toward a proper
seismic diagnostic for rotation.

Observing rotational splittings at lower frequencies than reported
by \cite{2012A&A...540A.143M} is challenging. First, such an
observation requires the presence of gravity-dominated mixed
modes. At low $\numax$, such modes have so long lifetimes that
very long observations are necessary to unveil them. Second,
disentangling mixed-mode spacing and rotational splitting at low
$\numax$ is demanding, especially for RGB stars. We shall
consider, as an example, an RGB star that has $\Dnu=4\,\mu$Hz and
$\numax = 35\,\mu$Hz; such a star has to be distinguished from its
more evolved counterpart in the red clump, which has similar
$\Dnu$ and $\numax$. Such a RGB star has a period spacing $\Tg
\simeq 50\pm3$\,s. The extrapolation of the measurements in
\cite{2012A&A...540A.143M} provides a rotational splitting
$\delta\nu\ind{rot}$ in the range [200 -- 300\,nHz]. At $\numax$,
the frequency spacing associated to $\Tg$ is of the order of
70\,nHz. This means that the multiplet components of 6 to 9 radial
orders are superimposed in a same small frequency range.
Disentangling them will be highly difficult, if not impossible.

\section{Conclusion}

At the time this paper is written, red giant seismology is still
in progress. It is however clear that we have efficient tools for
analyzing the spectra. The universal red giant oscillation pattern
has proven to provide the most precise measurement of the large
separation; furthermore, it provides the full identification of
the modes, i.e. their radial order and angular degree. The
asymptotic expansion for mixed modes has proven to precisely
depict the dipole mixed-mode pattern. The observation of large
gravity radial orders ensures a precise measurement of the
gravity-mode spacing. The toy model describing the coupling of
pressure and gravity waves allows us to better understand the
mixed modes.

The efficient work done by the CoRoT and \Kepler\ red giant
working groups ensures a permanent flux of new discoveries.
Intensive work is in progress, for a thorough understanding of
global and individual properties of red giants: ensemble
seismology, interior structure, stellar physics, astrophysical
consequences for stellar population or Galactic physics.

With dipole mixed modes, we have an efficient way to extract
information from the stellar core. With gravity period spacings,
we can explore the physics of the stellar core, including the
nuclear reactions at the different evolutionary stages. With the
rotational splittings, we can explore the inner angular momentum
and its evolution in red giants. With the observation of depressed
mixed modes, we can identify and study stars with very close
Brunt-V\"ais\"al\"a and Lamb cavities.

The accuracy of scaling relations providing the stellar mass and
radius is enhanced by all the calibration work. This calibration
can be done with theoretical work (Belkacem et al.~2011), with
specific studies of cluster stars (Miglio et al.~2012), or by
modelling (Belkacem et al.~2013). This modelling work shows that,
following \cite{2013A&A...550A.126M}, it is necessary to
distinguish between the large separation observed at $\numax$, the
asymptotic value, and the dynamical frequency which scales as
$\sqrt{M/R^3}$.

All this study relied on a set of about 1400 red giants studied
since \cite{2010ApJ...713L.176B}. The amount of data will be
increased by a factor of twenty with CoRoT data and \Kepler\
public data (Stello et al.~2013).

\begin{acknowledgements}

The authors acknowledge financial support from the ``Programme
National de Physique Stellaire" (PNPS, INSU, France) of CNRS/INSU
and from the ANR program IDEE ``Interaction Des Etoiles et des
Exoplan\`etes'' (Agence Nationale de la Recherche, France).

\end{acknowledgements}


\end{document}